\documentclass[12pt,twoside]{article}
\usepackage[mathscr]{eucal}
\usepackage{amsmath,amsfonts,amssymb,amsthm}
\usepackage[matrix,arrow]{xy}
\usepackage{times}
\usepackage{pdfsync}
\usepackage{graphicx}% Include figure files
\usepackage{epstopdf}
\usepackage{multibox}
\usepackage{dcolumn}
\usepackage{hyperref}

\def\d_Vphi{\text{d}_V\hspace{-0.06em}\phi}
\def\d_Vphibar{\text{d}_V\hspace{-0.06em}\bar\phi}
\def\d_Vxi{\text{d}_V\hspace{-0.06em}\xi}

\def\be{\begin{eqnarray}}
\def\ee{\end{eqnarray}}
\def\beann{\begin{eqnarray*}}
\def\eeann{\end{eqnarray*}}
\def\beq{\begin{equation}}
\def\eeq{\end{equation}}
\def\ba{\begin{array}}
\def\ea{\end{array}}
\def\ben{\begin{enumerate}}
\def\een{\end{enumerate}}
\def\bea{\begin{eqnarray}}
\def\eea{\end{eqnarray}}

\def\5{\bar }
\def\6{\partial }
\def\7{\hat }
\def\4{\tilde }
\advance\textheight\topskip
\renewcommand{\tilde}{\widetilde}
\renewcommand{\hat}{\widehat}
\newtheorem{prop}{Proposition}[section]

\renewcommand{\simeq}{\cong}

\newcommand{\dd}{\partial}
\renewcommand{\d}{\partial}

\renewcommand{\geq}{\,{\geqslant}\,}
\renewcommand{\leq}{\,{\leqslant}\,}

\newcommand{\binner}[2]{%
  {\langle}\kern-4.15pt{\langle}#1{,}\,#2{\rangle}\kern-4.15pt{\rangle}}

\newcommand{\half}{\frac{1}{2}}

\newcommand{\ffrac}[2]{\raisebox{.5pt}%
  {\footnotesize$\displaystyle\frac{#1}{#2}$}\kern1pt}

\newcommand{\dover}[2]{\ffrac{\dd #1}{\dd #2}}

\newcommand{\ddl}[2]{\ffrac{\dd #1}{\dd #2}}

\newcommand{\vddr}[2]{\ffrac{\delta^R #1}{\delta #2}}
\newcommand{\vddl}[2]{{\ffrac{\delta #1}{\delta #2}}}
\newcommand{\vddll}[2]{{\ffrac{\delta^L #1}{\delta #2}}}

\def\cA{\mathcal{A}}
\def\cB{\mathcal{B}}

\def\cF{\mathcal{F}}

\def\cJ{\mathcal{J}}

\def\cL{\mathcal{L}}

\numberwithin{equation}{section} \makeatletter
\@addtoreset{equation}{section}

\begin{document}

\def\mytitle{A note on gauge systems from the point of view of Lie
  algebroids}

\pagestyle{myheadings} \markboth{\textsc{\small Barnich}}{%
  \textsc{\small Gauge algebroid}}
\addtolength{\headsep}{4pt}

\begin{flushright}\small
ULB-TH/10-31, ESI-preprint 2265
\end{flushright}

\begin{centering}

  \vspace{1cm}

  \textbf{\Large{\mytitle}}

%\vspace{1cm}

%{\huge Notes}

  \vspace{1.5cm}

  {\large Glenn Barnich$^{a}$}

\vspace{.5cm}

\begin{minipage}{.9\textwidth}\small \it \begin{center}
   Physique Th\'eorique et Math\'ematique\\ Universit\'e Libre de
   Bruxelles\\ and \\ International Solvay Institutes \\ Campus
   Plaine C.P. 231, B-1050 Bruxelles, Belgium \end{center}
\end{minipage}

\end{centering}

\vspace{1cm}

\begin{minipage}{.9\textwidth}
  Proceedings for the {\bf XXIX Workshop on Geometric Methods in
    Physics}, 27.06-03.07.2010, Bia{\l}owie\.{z}a, Poland
  \end{minipage}

\vspace{1cm}

\begin{center}
  \begin{minipage}{.9\textwidth}
    \textsc{Abstract}.  In the context of the variational bi-complex,
    we re-explain that irreducible gauge systems define a particular
    example of a Lie algebroid. This is used to review some recent and
    not so recent results on gauge, global and asymptotic symmetries.
  \end{minipage}
\end{center}

\vfill

\noindent
\mbox{}
\raisebox{-3\baselineskip}{%
  \parbox{\textwidth}{\mbox{}\hrulefill\\[-4pt]}}
{\scriptsize$^a$Research Director of the Fund for Scientific
  Research-FNRS. E-mail: gbarnich@ulb.ac.be}

% \classification{02.20.Tw, 02.40.Hw, 02.40.Ky, 04.20.Fy,
%   04.20.Ha, 11.10.Ef} 
% \keywords {Gauge symmetries, global symmetries,
%   asymptotic symmetries, formal variational
%   calculus, Batalin-Vilkovisky formalism, local BRST cohomology, Lie
%   algebroids}

\thispagestyle{empty}
\newpage

% \begin{small}
% {\addtolength{\parskip}{-1.5pt}
%  \tableofcontents}
% \end{small}
% \newpage

\section{Introduction}

Gauge systems feature prominently in theoretical physics because the
four known fundamental interactions, electromagnetism, the weak and
strong nuclear forces, general relativity, and various unifying models
such as string or higher spin theories, are described by theories of
this type. It is therefore of interest to study the mathematical
structure of such systems.

More concretely, by gauge systems we mean systems of under-determined
partial differential equations deriving from variational
principles. In a first approximation, one often replaces the fields,
i.e., the dependent variables, by coordinates $\phi^i$ on some finite
dimensional manifold and forgets about the independent variables. For
instance, the action functional then reduces to an ordinary function
$S_0(\phi^i)$.

When applied to such a finite-dimensional toy model, the algebraic
structure underlying the Batalin-Vilkovisky (BV) construction as
reviewed for instance in \cite{Henneaux:1992ig} involves formulas that
are reminiscent of those that occur in the context of Lie
algebroids. The general picture is well-known: the base space is the
space of solutions to the Euler-Lagrange equations, the algebra is the
algebra of field dependent gauge parameters, their image under the
anchor are the gauge symmetries; the latter form an integrable
distribution and partition solution space by gauge orbits.
More precisely, let us denote by $R^i_\alpha\d/\d\phi^i$ an
irreducible generating set of gauge symmetries, i.e., a set of vector
fields such that
\begin{equation*}
  R^i_\alpha\dover{S_0}{\phi^i}=0\,,\qquad
  N^i\dover{S_0}{\phi^i}=0\Longrightarrow N^i\approx f^\alpha
  R^i_{\alpha}\,,\label{eq:1} 
\end{equation*}
for some functions $f^\alpha$. We use Dirac's notation for a function
that vanishes when pulled back to the surface $\Sigma$ defined by ${\d
  S_0}/{\d\phi^i}=0$, $g\approx 0$, and say that $g$ vanishes weakly
or vanishes on-shell. It then follows that the vector fields
$R^i_\alpha\d/\d\phi^i$ are in involution on-shell. Furthermore,
on-shell, they determine structure functions and an associated Lie
algebroid involving the algebra of field dependent gauge parameters
$f^\alpha$ and the anchor $ f^\alpha R^i_\alpha\d/\d\phi^i$. In
particular for instance, the associated ``longitudinal'' differential
$\gamma$ coincides, up to notation, with the differential occurring in
the local description of a Lie algebroid as reviewed for instance in
section 2.1 of \cite{2006math.....11259L}.

The remaining part of the BV construction consists in getting an
off-shell description of this differential by using a Koszul-Tate
resolution with additional generators, the antifields. In the
variational case, the off-shell differential can then be shown to be
canonically generated through a generator $S$ in terms of a suitable
antibracket.

What makes the finite-dimensional toy model uninteresting per se, at
least locally, is that under standard regularity assumption one can
choose local coordinates that trivialize the whole construction. This
is the content of the abelianization theorem.

The formal extension to field theories proceeds by assuming that the
index $i$ includes the independent variables $x^\mu$ and, at the same
time, summations over $i$ include integrations over $x^\mu$, this is
the DeWitt notation, see e.g.~\cite{DeWitt:2003pm}. The
danger of this approach is that one easily forgets about derivatives,
and it is precisely the derivatives that make the whole construction
non trivial, even when working in a local coordinate system.

In the present note, we re-explain how irreducible gauge field
theories define a particular Lie algebroid. For concreteness, we
choose in this note to control the functional aspects of the problem
by working in the framework of the variational bi-complex. The last
part of the note is devoted to a summary of results that we have
derived in this context.

Other approaches realizing the general picture are of course also
possible. In particular, in the context of asymptotic symmetries one
deals with concrete subspaces of solutions determined by some fall-off
conditions. In the conclusion, we re-interpret some of our results on
asymptotic symmetries from the perspective of Lie algebroids.

\section{Generalities}
\label{sec:vari-eom-symm}

In this section, we quickly review basic definitions and results on an
algebraic approach to symmetries. More details and proofs can be found
for instance
in~\cite{Andersonbook,Anderson1991,Dickey:1991xa,Olver:1993} and
references therein.

\subsection{Jet-bundles and Euler-Lagrange derivatives}
\label{sec:jet-bundles-euler}

Consider a fiber bundle $E$ with base space $M$. In the following, we
restrict ourselves to local coordinates $x^\mu$ on $M$ and $\phi^i$ on
the fiber $C$. Coordinates on the associated jet-bundle $\cJ^k$ of
order $k$ are denoted by $x^\mu,\phi^i_{(\mu)}$. Here $(\mu)$ stands
for an unordered index $\mu_1\dots\mu_l$, with $l\leq k$. For such an
index, $|\mu|=l$. The total derivative is the operator
\begin{eqnarray}
  \label{eq:29}
  \d_\nu=\ddl{}{x^\nu}+\phi^i_{((\mu)\nu)}\ddl{}{\phi^i_{(\mu)}},
\end{eqnarray}
where the summation conventions for repeated indices is used.  A local
function is a smooth functions on $\cJ^k$ for some finite $k$. The
space of local functions is denoted by $Loc(E)$.

If $(-\d)_{(\mu)}=(-)^{|\mu|}\d_{(\mu)}$, the Euler-Lagrange
derivative of a local function $f$ is defined by
\begin{eqnarray}
  \label{eq:30}
  \vddl{f}{\phi^i}=(-\d)_{(\mu)}\ddl{f}{\phi^i_{(\mu)}}.
\end{eqnarray}
The adjoint of a total differential operator
$O=O^{(\mu)} \partial_{(\mu)}$ is $O^\dagger\cdot
=(-\d)_{(\mu)}(O^{(\mu)}\cdot)$ so that $(O^\dagger)^\dagger=O$.  For
a collection of local functions $P^a$, the Fr\'echet derivative is the
matrix-valued total differential operator defined by
\begin{equation}
{{D_P}^a}_j=\dover{P^a}{\phi^j_{(\nu)}}\d_{(\nu)}\label{eq:64}\,.
\end{equation}
Note also that the Fr\'echet derivative can be
defined for a collection of total differential operators
$O^a=O^{a(\mu)}\d_{(\mu)}$ through ${{D_O}^a}_j\equiv
{{D_{O^{(\mu)}}}^a}_j\circ \d_{(\mu)}$. 

\subsection{Stationary surface}
\label{sec:stationary-surface}

Equations of motion are partial differential equations of the form
$E_a[\phi]=0$ where $E_a$ are local functions that vanish when the
fields and their derivatives are put to zero. The equations of motion
$E_a[\phi]=0$ are variational if the range of $a$ and $i$ are the same
and if there exists a local function $L$ called Lagrangian such that
\begin{eqnarray}
E_i=\vddl{L}{\phi^i}\label{eq:31}.
\end{eqnarray}
This is the case if and only if 
\begin{eqnarray}
  \label{eq:14}
  {D_E}_{ij}=({D_E}_{ji})^\dagger. 
\end{eqnarray}

The ``stationary'' surface $\Sigma$ is defined in the jet-bundles by
the equations of motion and their total derivatives,
\begin{eqnarray}
  \label{eq:33}
  \d_{(\mu)}E_a=0.
\end{eqnarray}
Under appropriate regularity conditions (see
e.g.~\cite{Henneaux:1992ig}) which we assume to be fulfilled,
$f\approx 0$ if and only if there exists local functions $k^{a(\mu)}$
such that $f=k^{a(\mu)}\d_{(\mu)}E_a$. The space $Loc(\Sigma)$ of
local functions on $\Sigma$ can then be identified with $Loc(E)/I$
where $I$ is the ideal of local functions vanishing on $\Sigma$. The
associated space of local forms on $\Sigma$ is denoted by
$\Omega_\Sigma$. 

\subsection{Horizontal complex and prolongation of generalized
  vector fields}
\label{sec:horizontal-complex}

The horizontal complex is the Grassmann algebra generated by the odd
elements $dx^\mu$ with coefficients that are local functions,
$\Omega=Loc(E)\otimes \wedge (dx^\mu)$. The horizontal differential is
$d_H=dx^\mu\d_\mu$. A generalized vector field is a vector field of
the form $X=P^\mu\ddl{}{x^\mu}+R^i\ddl{}{\phi^i}$, with $P^\mu,R^i$
local functions. Its prolongation to horizontal forms is defined by
\begin{equation}
  \label{eq:8}
  pr\,
  X=\d_{(\mu)}Q^i\ddl{}{\phi^i_{(\mu)}}+P^\mu\d_\mu+d_HP^\mu\ddl{}{dx^\mu},\quad Q^i=R^i-P^\mu\d_\mu\phi^i,
\end{equation}
in such a way as to commute with the horizontal differential 
\begin{eqnarray}
  \label{eq:32}
   [pr\, X,d_H]=0.
\end{eqnarray}
The horizontal complex pulled back to the stationary surface is
denoted by $\Omega_\Sigma$. 

\subsection{Local functionals}
\label{sec:local-functionals}

The space of local functionals $\cF$ is defined by
$\cF=H^n(d_H,\Omega)$. A local functional is thus an equivalence class
$[\cL]$, $\cL=Ld^nx$ where $L\sim L+\d_\mu k^\mu$, with $L,k^\mu$
local functions, i.e., a Lagrangian $L$ up to a total divergence.  The
property
\begin{equation}
  \label{eq:52}
  \vddl{L}{\phi^i}=0\iff L=\d_\mu k^\mu,
\end{equation}
allows one to characterize local functionals as equivalence classes of
Lagrangians with identical Euler-Lagrange derivatives. The action is
the distinguished local functional $S_0=[\cL]$ whose associated
Euler-Lagrange derivatives define the equations of motion.

\subsection{Equations of motion and variational symmetries}
\label{sec:equat-moti-vari}

A generalized vector field $X$ defines an equations of motion symmetry
if 
\begin{equation}
pr\, X E_a \approx 0\label{eq:46}.
\end{equation}
A generalized vector field $X$ defines a variational symmetry of the
action $[\cL]$ if
\begin{eqnarray}
  \label{eq:10}
  pr\, X \cL =d_H k. 
\end{eqnarray}
If $Q^i=0$, $X$ is both an equations of motion and a variational
symmetry for all $P^\mu$. We will thus restrict ourselves in the
following to generalized vector fields in evolutionary form,
$Q=Q^i\ddl{}{\phi^i}$, with prolongation
\begin{eqnarray}
  \label{eq:9}
  \delta_Q=\d_{(\mu)}Q^i\ddl{}{\phi^i_{(\mu)}}.
\end{eqnarray}
The following formulae which can be derived for instance from
Eq.~(6.42) and Eq.~(6.43) of \cite{Barnich:2000zw}, are useful in the
following:
\begin{align}
  \label{eq:65}
  [\delta_Q,\vddl{}{\phi^j}]&=-({{D_Q}^{i}}_j)^\dagger\circ
  \vddl{}{\phi^i},\\
\delta_{Q_1}
({{D_{Q_2}}^{i}}_j)^\dagger&=({{D_{\delta_{Q_1}Q_2}}^{i}}_j)^\dagger
-({{D_{Q_2}}^i}_k\circ {{D_{Q_1}}^k}_j)^\dagger.\label{eq:65b}
\end{align}

By applying an Euler-Lagrange derivative to $\delta_Q L=\d_\mu k^\mu$,
an evolutionary vector field defines a variational symmetry if and
only if 
\begin{eqnarray}
  \label{eq:11}
  \delta_Q \vddl{L}{\phi^j}=-({{D_Q}^{i}}_j)^\dagger[\vddl{L}{\phi^i}].
\end{eqnarray}
It follows that every variational symmetry is an equations of motion
symmetry. 

Evolutionary vector fields ($EV$), equations of motion symmetries
($MS$) and variational symmetries ($VS$) are Lie algebras with bracket
\begin{equation}
  \label{eq:53}
  [Q_1,Q_2]^i=\delta_{Q_1} Q^i_2 -\delta_{Q_2} Q^i_1,\quad 
  [\delta_{Q_1},\delta_{Q_2}]=\delta_{[Q_1,Q_2]}\,.
\end{equation}

\subsection{On-shell symmetries}
\label{sec:gauge-symmetries}

Evolutionary vector fields such that $Q^i\approx 0$ define equations
of motion symmetries. Such equations of motion symmetries are
considered trivial. They form a Lie ideal. Proper equations of motion
symmetries are defined as equivalence classes of equations of motion
symmetries modulo trivial ones. They restrict to well defined vector
fields on $\Sigma$. We denote the Lie algebra of proper equations of
motion symmetries by $PMS$. 

Similarly, variational symmetries such that $Q^i\approx 0$ form an
ideal in the Lie algebra of variational symmetries.

\subsection{Generalized conservation laws}
\label{sec:gener-cons-laws}

Generalized conservation laws correspond to the cohomology
spaces $H^{n-k}(d_H,\Omega_\Sigma)$ with $k\geq 1$ defined by 
\begin{eqnarray}
  \label{eq:34}
  H^{n-k}\big(d_H,\Omega_\Sigma\big)\ni[\omega^{n-k}]
  \iff\left\{\begin{array}{l} 
  d_H\omega^{n-k}\approx 0,\\ \omega^{n-k}\sim \omega^{n-k}
  +d_H\eta^{n-k-1}+t^{n-k},\ t^{n-k}\approx 0. 
\end{array}\right.
\end{eqnarray}

\section{Gauge and global symmetries}
\label{sec:bv-description}

\subsection{Noether operators}
\label{sec:noether-operators}

A Noether operator is a total differential operator $N^a\equiv
N^{a(\mu)}\d_{(\mu)}$ such that
\begin{equation}
  \label{eq:41}
  N^{a}[E_a]=0.
\end{equation}
The linear space of Noether operators (NO) is a left module over the
associative algebra of total differential operators.

A set of Noether operators $R^{\dagger
  a}_{\alpha}$ is a generating set\footnote{To agree with standard usage,
  the generating set is usually expressed in terms of adjoints of some
  operators $R^a_\alpha$. } if every Noether operator $N^a$ can be
written in terms of the generating set on-shell, i.e., if there exists
operators $O^{\alpha}\equiv O^{\alpha(\mu)}\d_{(\mu)}$ such that
\begin{equation}
  \label{eq:42}
  N^a\approx O^{\alpha}\circ R^{\dagger a}_{\alpha}.
\end{equation}
We assume here for simplicity of
the arguments below that the generating set is irreducible, i.e., that
for all operators $Z^\alpha$, 
\begin{equation}
  \label{eq:43}
  Z^{\alpha}\circ R^{\dagger a}_{\alpha}\approx 0\Longrightarrow
  Z^{\alpha}\approx 0. 
\end{equation}

In the rest of this section, we concentrate on the case of variational
equations associated with an action $S_0=[\cL]$. The associative
algebra $TDO$ of total differential operators is a Lie module over
$VS$ under the action of $\delta_Q$ with the Leibniz rule  
\begin{equation}
  \label{eq:55}
  \delta_{Q} (O_1\circ O_2)=\delta_{Q} O_1\circ
  O_2+O_1\circ\delta_{Q} O_2.
\end{equation}
\begin{prop}
  Noether operators are a module over $VS$,
  \begin{equation}
    \label{eq:63}
    (Q\cdot N)^i=\delta_Q N^i-N^j\circ ({{D_Q}^i}_j)^\dagger. 
  \end{equation}
\end{prop}
{\bf Proof:} Applying a
  variational symmetry to a Noether identity gives
  \begin{equation*}
0= \delta_Q \Big(N^{i}[\vddl{L_0}{\phi^i}]\Big)=\delta_Q
( N^{i(\mu)})\partial_{(\mu)} \vddl{L_0}{\phi^i}-
(N^i\circ ({{D_Q}^j}_i)^\dagger)[\vddl{L_0}{\phi^j}],
\end{equation*}
by using (\ref{eq:11}). This implies that the RHS of \eqref{eq:63} is
a Noether operator. That 
\begin{equation}
Q_1\cdot (Q_2\cdot N)-Q_2\cdot (Q_1\cdot
N)=[Q_1,Q_2]\cdot N\label{eq:69}
\end{equation}
follows from a straightforward computation using
\eqref{eq:65b}.\qed

It also follows directly from \eqref{eq:63} that 
\begin{equation}
  \label{eq:67}
  Q\cdot (O\circ N)=(\delta_Q O)\circ N + O\circ (Q\cdot N). 
\end{equation}

\subsection{Gauge symmetries}
\label{sec:gauge-glob-symm}

Standard integrations by parts show that there is a linear map $\rho$
from the space of Noether operators to the space of variational
symmetries: if $N^i$ is a Noether operator, the characteristic of the
associated variational symmetry is $\rho(N)^i=N^{\dagger i} (1)$. Note
in particular that $\rho(N\circ D^\dagger_Q)=\delta_{\rho(N)}Q$.

The space of gauge symmetries $GS$ is defined as the subspace
$\text{Im}\ \rho\subset VS$. It is a Lie ideal in the space of variational
symmetries. This follows from the crucial property
\begin{equation}
  \label{eq:66}
\rho(Q \cdot N)= [Q,\rho(N)]\,. 
\end{equation}
Another property of $\rho$ which can be
proved by using again formula  Eq.~(6.43) of \cite{Barnich:2000zw} is 
\begin{equation}
  \label{eq:2}
  D^\dagger_{\rho(N)}=D^\dagger_N. 
\end{equation}
One then can use $\rho$ to define a bilinear map on Noether operators
through $N_1\star N_2=\rho(N_1)\cdot N_2$. Even though this map is not
skew-symmetric, its image under $\rho$ is due to
\eqref{eq:66}. Furthermore, as a consequence of \eqref{eq:69}, it
satisfies $N_1\star (N_2\star N_3)-N_2\star(N_1\star N_3)-(N_1\star
N_2)\star N_3=0$ which is mapped to the Jacobi identity  for gauge
symmetries when applying $\rho$.

\subsection{Global symmetries}
\label{sec:global-symmetries}

By definition, the quotient Lie algebra $VS/GS$ of variational
symmetries modulo gauge symmetries is the Lie algebra of global
symmetries. 

\subsection{Proper gauge symmetries}
\label{sec:prop-gauge-symm}

Trivial total differential operators or Noether operators are defined
by operators whose coefficients vanish on-shell. Multiplication of a
Noether operator by a trivial operator gives a trivial Noether
operator. Trivial gauge symmetries are variational symmetries that lie
in the image of trivial Noether operators. They form an ideal in the
Lie algebra of gauge symmetries. Proper total differential operators,
Noether operators, gauge symmetries are defined as total differential
operators, Noether operators, gauge symmetries modulo trivial
ones. 

\subsection{Gauge algebroid}
\label{sec:gauge-algebroid}

Proper gauge symmetries are generated by $\rho(O^\alpha\circ
R^{i\dagger}_\alpha)$ with the equivalence relation $O^\alpha\sim
O^\alpha+t^\alpha$ and where $TDO\ni t^\alpha\approx 0$. Let us
introduce the notations $\rho(O^\alpha\circ
R^{i\dagger}_\alpha)=R^{i}_{\alpha}(f^\alpha)= R^i_f$ where
$f^\alpha=O^{\dagger\alpha}(1)$, and also
$\delta_{f}=\delta_{R_f}$. Proper gauge symmetries are thus also
generated by variational symmetries with characteristic
$R^i_\alpha(f^\alpha)$ where $f^\alpha\in Loc(\Sigma)$. Note that
irreducibility of $R^{i\dagger}_\alpha$ can easily be shown to be
equivalent to the statement that if $R^i_\alpha(O^\alpha(g))\approx 0$
for all $g\in Loc(E)$ then $O^\alpha\approx 0$. The property that
$R^{i\dagger}_\alpha$ is a generating set is equivalent to the
statement that any family of variational symmetries that depends
linearly and homogeneously on an arbitrary local function $f$ and its
derivatives, $G^i(f)=G^{i(\mu)}\d_{(\mu)} f$ and $\delta_G L=\d_\mu
k^\mu(f)$ can be written as $G^i(f)\approx R^i_\alpha(O^\alpha(f))$
for some $O^\alpha\in TDO$.

Since $[R_{f_1},R_{f_2}]$ defines a variational symmetry, one can
easily prove from the generating property that
\begin{equation}
  \label{eq:5}
  [R_{f_1},R_{f_2}]^i\approx
  R^i_\gamma\big(C^\gamma_{\alpha\beta}(f^\alpha_1,f^\beta_2)+
\delta_{f_1}f^\gamma_2-\delta_{f_2}f^\gamma_1\big)\,, 
\end{equation}
where $C^\gamma_{\alpha\beta}(f^\alpha_1,f^\beta_2)=
C^{\gamma(\mu)(\nu)}_{\alpha\beta}\d_{(\mu)}f_1^\alpha\d_{(\nu)}
f_2^\beta $ are bi-differential operators that are skew-symmetric,
$C^\gamma_{\alpha\beta}(f^\alpha_1,f^\beta_2)=-
C^\gamma_{\beta\alpha}(f^\beta_2,f^\alpha_1)$. Introducing a linear
space spanned by $e_\alpha$ associated with the generating set of
Noether operators $R^{\dagger i}_\alpha$ and defining $A$ as the
linear space with elements $f=f^\alpha e_\alpha$ where $f^\alpha\in
Loc(\Sigma)$, $A$ is a Lie algebra with bracket
\begin{equation}
  \label{eq:6}
  [f_1,f_2]_A=\big(C^\gamma_{\alpha\beta}(f^\alpha_1,f^\beta_2)+
\delta_{f_1}f^\gamma_2-\delta_{f_2}f^\gamma_1)e_\gamma\,. 
\end{equation}
Indeed, the Jacobi identity for the bracket $[\cdot,\cdot]_A$ is a
direct consequence of the Jacobi identity for the bracket of
evolutionary vector fields applied to $R_{f_1},R_{f_2},R_{f_3}$ and
the irreducibility of the generating set. 

To an irreducible gauge theory and a choice of generating set
$R^{\dagger i}_\alpha$, one can thus associate the Lie algebroid with
algebra $A$ as a vector bundle over the stationary surface $\Sigma$
with anchor the map $a(f)=\delta_f$. For want of a better name, one
may call this the gauge algebroid. 

Up to details related to the treatment in the context of the
variational bi-complex, there is of course no claim of
originality. Indeed, in some way or the other, this is known to most
people familiar with the Batalin-Vilkovisky construction, see for
instance \cite{Lyakhovich:2004kr}. Related considerations have
appeared for instance in \cite{Fulp:2002kk}. Note that the off-shell
description gives rise to an sh-Lie algebroid, while $L$-stage
reducible gauge theories are $L$-Lie algebroids. This is most
transparent in the antifield formalism to which we now turn. 
 
\section{BV description}
\label{sec:bv-description-1}

Both in the variational and the non-variational case, a description
with antifields and ghosts originating from the Batalin-Vilkovisky
approach
\cite{Batalin:1981jr,Batalin:1983wj,Batalin:1983jr,Batalin:1984ss,%
  Batalin:1985qj} to the quantization of Lagrangian gauge field
theories turns out to be useful.

Various elements of the construction appear in
\cite{Henneaux:1985kr,Henneaux:1988ej,Fisch:1989dq,Fisch:1990rp} and
are summarized in \cite{Henneaux:1992ig}. The non-variational case has
been studied in \cite{Henneaux:1989cz}. Aspects related to locality
and jet-bundles are treated in
\cite{Henneaux:1991rx,Barnich:1994db,Barnich:1994mt,Barnich:1996mr,%
Barnich:2000zw,Barnich:2001jy}.

\subsection{Homological resolution of on-shell functions}
\label{sec:homol-resol-shell}

For an irreducible set of Noether operators, the fiber is extended to
include the ``anti-fields'' $\phi^*_a$ (even) and $C^*_{\alpha}$ (odd),
of resolution degrees $1$ and $2$ respectively with all other
variables of degree $0$. We denote the space of local functions on
this extended space by $Loc(E^{AF})$. The homology of the
(evolutionary) homological vector field
\begin{equation}
  \label{eq:44}
  \delta=\d_{(\mu)} 
R^{\dagger a}_{\alpha}[\phi^*_a]\dover{}{C^*_{\alpha(\mu)}}+ \d_{(\mu)} E_a
\dover{}{\phi^*_{a(\mu)}},\qquad \delta^2=0,
\end{equation}
is 
\begin{equation}
  \label{eq:45}
  H_k(\delta,Loc(E^{AF}))=\begin{cases} 0\ &\text{for}\ k>0\\
C^\infty(\Sigma)\ &\text{for}\ k=0.
\end{cases}
\end{equation}
It follows that 
\begin{prop}
  The Lie algebra $PMS$ of proper equations of motion symmetries is
  isomorphic to $H_0([\delta,\cdot])$, the adjoint cohomology of
  $\delta$ in the space of evolutionary vector fields acting on
  $Loc(E^{AF})$ in resolution degree $0$ equipped with the induced Lie
  bracket for evolutionary vector fields.
\end{prop}
Furthermore, $H_k([\delta,\cdot],EV_{E^{AF}})=0, k\geq 1$. 

\subsection{Longitudinal differential}
\label{sec:long-diff}

Consider a subset of equations of motion symmetries
$\delta_A$ with  characteristic $Q_A$ that are integrable on-shell,
\begin{equation}
  \label{eq:47}
  [\delta_A,\delta_B]\approx f^C_{AB}\delta_C\,,
\end{equation}
where $f^A_{B C}$ are local functions. 

Consider the pure ghost number, i.e., the degree for which $C^A$ are
Grassmann odd generators of degree $1$ with all other variables in
degree $0$.  On the space $Loc(\Sigma)\otimes \wedge (C^A)$, the
associated homological vector field (``longitudinal differential'') is
\begin{equation}
  \label{eq:49}
  \gamma=C^A\delta_A -\half C^A C^B  f^C_{AB} \dover{}{C^C}, \qquad
  \gamma^2\approx 0. 
\end{equation}

\subsection{Homological perturbation theory}
\label{sec:homol-pert-theory}

Consider the space $Loc(E^{AF})\otimes \wedge (C^A)$ with total degree
(``ghost number'') the pure ghost number minus the resolution
degree. The main theorem on the off-shell description of the
longitudinal differential and its cohomology says that perturbatively
in the resolution degree, there exists a differential $s$ (``BRST
differential'') on this space
\begin{equation}
  \label{eq:50}
  s=\delta+\gamma+s_1+\dots, \qquad s^2=0,
\end{equation}
such that 
\begin{equation}
  \label{eq:51}
  H^k(s,Loc(E^{AF})\otimes \wedge (C^A))=\begin{cases} 0\ &\text{for}\ k<0\\
H^k(\gamma,Loc(\Sigma \otimes \wedge (C^A))\ &\text{for} \ k\geq 0.
\end{cases}
\end{equation}

\subsection{Longitudinal differential for proper gauge symmetries}
\label{sec:long-diff-prop}

For proper gauge symmetries associated to the generating set
$R^{i\dagger}_\alpha$ we extend the fiber by odd generators $C^\alpha$
``ghosts'' and the associated longitudinal differential can be written
as
\begin{equation}
  \gamma=\d_{(\rho)}(R^{i}_\alpha(
  C^\alpha))\ddl{}{\phi^i_{(\rho)}}
  -\half\d_{(\rho)}(C^\gamma_{\alpha\beta}(C^\alpha,C^\beta))
\ddl{}{C^\gamma_{(\rho)}},
\label{eq:71}
\end{equation}
with
$C^\gamma_{\alpha\beta}(f^\alpha_1,f^\beta_2)=C^{\gamma(\mu)(\nu)}_{\alpha\beta}\d_{(\mu)}f^\alpha_1\d_{(\nu)}f^\beta_2$
total bi-differential skew-symmetric operators. This differential is
of course just the standard Lie algebroid differential in the
particular case of the gauge algebroid.

\subsection{Master action}
\label{sec:master-action}

In the extended fiber with ghosts and antifields, $C^\alpha$ are of
ghost number $1$, $\phi^*_i$ of ghost number $-1$ and $C^*_\alpha$ of
ghost number $-2$. All other variables are of ghost number $0$.

Let $z^a=(\phi^i,C^\alpha)$. There is an odd graded Lie algebra
structure ``antibracket'' on the space of local functionals
$[\cA=ad^nx]$ defined by
\begin{gather}
  \label{eq:72}
(\cdot,\cdot):\cF^{g_1}\times\cF^{g_2}\rightarrow
\cF^{g_1+g_2+1},\nonumber
\\
  ([\cA_1],[\cA_2])=\Big[\big(\vddr{a_1}{z^a}\vddll{a_2}{z^*_a}-
(z\leftrightarrow z^*)\big)d^nx\Big].
\end{gather}
The evolutionary vector field associated with a functional $\cA$ is
then
\begin{equation}
(\cA,\cdot)_{alt}=\big(\d_{(\mu)}\vddr{a}{z^a}\dover{{}^L}{z^*_{a(\mu)}}-
(z\leftrightarrow z^*)\big).\label{eq:74}
\end{equation}

In the variational case, the BRST differential $s$ is canonically
generated by a master action $S$ of ghost number $0$, 
\begin{gather}
  \label{eq:73}
  s=(S,\cdot)_{alt},\quad \half (S,S)=0,\nonumber \\
S=\Big[\big(L+\phi^*_iR^i_\alpha(C^\alpha)+\half C^*_\gamma
f^\gamma_{\alpha\beta}(C^\alpha,C^\beta)+\dots\big)d^nx\Big].
\end{gather}

\subsection{ Local BRST cohomology}
\label{sec:local-brst-cohom}

The cohomology space $H^*(s,\cF)$ of the BRST differential in the
space of local functionals is an odd graded Lie algebra for the
antibracket induced in cohomology.  Under suitable assumptions, one
can prove the following results for irreducible gauge theories
considered here:
\begin{enumerate}
\item $H^{g}(s,\cF)\simeq H^{n+g}(d_H,\Omega_\Sigma)=0$ for $g\leq -3$.
\item $H^{-2}(s,\cF)\simeq H^{n-2}(d_H,\Omega_\Sigma)$ is isomorphic
  to the space of equivalence classes of reducibility parameters
  $[f^\alpha]$, where $f^\alpha$ are collections of local functions
  such that $R^i_\alpha(f^\alpha)\approx 0$ with $f^\alpha\sim
  f^\alpha +t^\alpha$ and where $t^\alpha\approx 0$. 
\item $H^{-1}(s,\cF)\simeq H^{n-1}(d_H,\Omega_\Sigma)$ is isomorphic
  to the space of global symmetries. 
\item Every variational symmetry with weakly vanishing characteristic
  is a gauge symmetry and thus trivial as a global symmetry. It
  follows that global symmetries are a sub-Lie algebra of proper
  equations of motion symmetries, $VS/GS\subset PMS$. 
\end{enumerate}
Furthermore, up to a suspension, the antibracket induced in
$H^{-1}(s,\cF)$ coincides with the Lie bracket of global
symmetries. The Lie bracket induced in the space of equivalence classes of
conserved currents $H^{n-1}(d_H,\Omega_\Sigma)$ is defined by
\begin{equation}
[[j_1],[j_2]]=[-\delta_{Q_1} j_2]\,,\label{eq:77}
\end{equation}
where $Q_1$ is the variational symmetry associated with
$j_1$. Together with item 3 above, this provides a complete and
generalized version of Noether's first theorem for irreducible gauge
theories.

More generally, via the antibracket induced in cohomology,
$H^{g}(s,\cF)$ is a module over the Lie algebra of global
symmetries. 

In addition, when $[S^{(1)}]\in H^0(s,\cF)$, there is a derived
(even) Lie bracket in $H^{-2}(s,\cF)$ defined by
\begin{equation}
[[\cA^{-2}],[\cB^{-2}]]=[(\cA^{-2},(S^{(1)},\cB^{-2}))].\label{eq:78}
\end{equation}
Through the isomorphism, it also induces a Lie algebra structure in
$H^{n-2}(d_H,\Omega_\Sigma)$.

\section{Discussion}
\label{sec:discussion}

From the definition of reducibility parameters in item 2 above and the
perspective of the present note, it follows that this space is
precisely the kernel of the anchor $a$. Furthermore, reducibility
parameters at a particular solution have also been considered. From
the point of view of Lie algebroids, they correspond to the isotropy
Lie algebra at a given point. They are related to the reducibility
parameters associated with the linearized gauge theory around this
solution. Together with the associated generalized conservation laws,
they have important physical applications. In gravity for instance,
they are the Killing vectors of the solution and the associated
conservation laws, also called surface charges, are related for
instance to the ADM energy-momentum. In the discussion of
integrability of these surface charges, paths in solution and gauge
parameter spaces have been considered
\cite{Barnich:2003xg,Barnich:2007bf}. It should prove most instructive
to try to understand better the relation to the Lie algebroid paths
and integrability discussed for instance in
\cite{2006math.....11259L}.

In the context of asymptotic symmetries, one does not work in the
framework of the variational bi-complex but one restricts oneself to
concrete and physically relevant subspaces of solutions. The claim is
the following:

{\em From the point of view of Lie algebroids, the results of
  \cite{Barnich:2010eb,Barnich:2009se} on asymptotically
  anti-de Sitter space-times in three dimensions at spatial infinity
  or asymptotically flat spacetimes in three or four dimensions at
  null infinity can be interpreted the sense that the associated gauge
  algebroid reduces to an action Lie algebroid for the Virasoro
  algebra in the former case and a suitable contraction or extension
  thereof in the latter two.}

\section*{Acknowledgments}
The author is grateful to the organizers of the XXIX Workshop on
Geometric Methods in Physics in Bia{\l}owie\.{z}a for the invitation
to present related work in a stimulating atmosphere. He thanks
Th. Voronov for a useful discussion on Lie algebroids and the modified
Lie bracket that arises in the context of asymptotic symmetries. The
author is Research Director of the Fund for Scientific Research-FNRS
(Belgium). This work is supported in part by the Belgian Federal
Science Policy Office through the Interuniversity Attraction Pole
P6/11, by IISN-Belgium, by ``
  Communaut\'e fran\c caise de Belgique - Actions de Recherche
  Concert\'ees'' and by Fondecyt Projects
No.~1085322 and No.~1090753.

% \bibliography{/Users/gbarnich/Documents/Literature/Bibliography/master2}

\def\cprime{$'$}
\providecommand{\href}[2]{#2}\begingroup\raggedright\endgroup

\end{document}